\documentclass[12pt,thmsa]{article}
\usepackage{sw20aip}

\input tcilatex
\QQQ{Language}{
American English
}

\begin{document}

\author{Freddy P. Zen$^{1)\thanks{$fpzen@bdg.centrin.net.id$}}\,\,$and \thinspace
Hendry I. Elim$^{1),2)\thanks{$hendry202@cyberlib.itb.ac.id$}}$ \and %
1)\thinspace Theoretical High Energy Physics Group, \and \thinspace
\thinspace \thinspace \thinspace \thinspace \thinspace Theoretical Physics
Lab.,\thinspace Department of Physics,\thinspace \and \thinspace \thinspace
\thinspace \thinspace \thinspace \thinspace \thinspace Institute of
Technology Bandung,\thinspace Bandung,\thinspace Indonesia \and 2)\thinspace
\thinspace Department of Physics,\thinspace Pattimura University, \and %
\thinspace \thinspace \thinspace \thinspace \thinspace \thinspace \thinspace
Ambon,\thinspace \thinspace Indonesia}
\title{Soliton Solution of the Integrable Coupled Nonlinear Schr\"{o}dinger
Equation of Manakov Type}
\date{December 23, 1998}
\maketitle

\begin{abstract}
The soliton solution of the integrable coupled nonlinear Schr\"{o}dinger
equation (\textbf{NLS}) of Manakov type\thinspace is\thinspace investigated
by using Zakharov-Shabat (\textbf{ZS}) scheme. We get the bright N-solitons
solution by solving the integrable uncoupled NLS of Manakov type. We also
find that \thinspace there is an elastic collision of the bright
N-solitons.\thinspace
\end{abstract}

\section{Introduction}

The integrable coupled nonlinear Schr\"{o}dinger equation of Manakov type is
widely used in recent developments in the field of optical solitons in
fibers. The use and applications of the equation is to explain how the
solitons waves transmit in optical fiber, what happens when the
interaction\thinspace among optical solitons influences directly the
capacity and quality of communication and so\thinspace on.$^{[1],[2],[3]}\,$%
On the other hand, many interesting physical phenomena can be modeled by
discrete nonlinear equations. Examples include vibration of particles in a
one-dimensional lattice, ladder type electric circuits, collapse of Langmuir
waves in plasma\thinspace physics, growth of conflicting populations in
biological science, different simulations of differential equation, etc.
Hence it is undoubtedly significant that the ideas of the inverse scattering
transform apply to certain types of discrete evolution equations.$^{[4]}$

Some of exact solutions have been derived for the system related to the
Manakov type equation: 
\begin{eqnarray}
iq_{1_x}+q_{1_{tt}}+2\mu \left( \left| q_1\right| ^2+\left| q_2\right|
^2\right) q_1 &=&0  \nonumber \\[0.01in]
iq_{2_x}+q_{2_{tt}}+2\mu \left( \left| q_1\right| ^2+\left| q_2\right|
^2\right) q_2 &=&0  \tag{1.1}
\end{eqnarray}
with different procedures $^{\left[ 5\right] ,\left[ 6\right] ,\left[
7\right] }\,$including the use of Hirota method$^{\left[ 8\right] }$. Based
on ref.[8], Radhakrishnan, \thinspace et. \thinspace al. got the one and two
soliton solution. However promising their methods, the solution of
N-solitons of the equation has not been derived in an exact result yet. In
this paper, we investigate the N-soliton solution. However, we show that it
is possible when the Zakharov-Shabat (\textbf{ZS}) scheme (in its final
form) is expressed solely in terms of certain operators. Based on the choice
of the operator, we find that there is a constraint$\,\,q_2=\sqrt{2m}%
\,q_1^{*}$ (where $m$ is an arbitrary positive real parameter and $q_1^{*}$
is a complex conjugate of $q_1$) when $\mu $ is an arbitrary positive real
value.

This paper is organized as follows. In section II, we will perform the 
\textbf{ZS} scheme and will obtain the constraint $q_2=\sqrt{2m}\,q_1^{*}$%
,\thinspace as well as the positive real parameter $\mu $. In section III,
we will solve the bright N-solitons solution. We will compare one and two
soliton solution with the result in ref.[$8$]. Section IV is devoted for
discussions and conclusions.

\section{ZS \thinspace Scheme\thinspace for\thinspace the\thinspace
\thinspace Integrable\thinspace \thinspace \thinspace Uncoupled \thinspace 
\textbf{NLS} of Manakov Type}

We start by choosing the following two operators related to Zakharov-Shabat (%
\textbf{ZS}) scheme 
\begin{equation}
\Delta _0^{(1)}\,=\,I\left( \frac i{\left( 2m+1\right) }\frac \partial
{\partial x}\,-\,\frac{\partial ^2}{\partial t^2}\right) ,  \tag{2.1a}
\end{equation}
$\,$and 
\begin{equation}
\Delta _0^{(2)}\,=\,\left( 
\begin{array}{cc}
m+1 & 0 \\ 
0 & -m
\end{array}
\right) \frac \partial {\partial t},  \tag{2.1b}
\end{equation}
where $m$ is an arbitrary positive real, and $I$ is the 2x2 unit matrix. We
can then define the following\thinspace operators by using this scheme
related to inverse scattering techniques$^{[2]}$%
\begin{equation}
\Delta ^{(1)}\,=\,\,\Delta _0^{(1)}+\,U\left( t,x\right) ,  \tag{2.2a}
\end{equation}
and 
\begin{equation}
\Delta ^{(2)}\,=\,\Delta _0^{(2)}+\,V\left( t,x\right) .  \tag{2.2b}
\end{equation}
Here operators $\Delta ^{(i)}$,($i\,$=\thinspace 1,\thinspace 2) satisfy the
following equation 
\begin{equation}
\Delta ^{(i)}\left( I\,+\,\Phi _{+}\right) \,=\,\left( I\,+\,\Phi
_{+}\right) \Delta _0^{(i)},  \tag{2.3}
\end{equation}
where the integral operator $\Phi _{\pm }\left( \psi \right) \,$\thinspace
are difined according to equation 
\begin{equation}
\Phi _{\pm }\left( \psi \right) \,=\,\int\limits_{-\infty }^\infty k_{\pm
}\left( t,z\right) \psi \left( z\right) dz.  \tag{2.4}
\end{equation}

We now\thinspace suppose that operators\thinspace $\Phi _F\left( \psi
\right) \,$and\thinspace \thinspace $\Phi _{\pm }\left( \psi \right) \,$ are
related to the following operator\thinspace identity 
\begin{equation}
\left( I\,+\,\Phi _{+}\right) \,\left( I\,+\,\Phi _F\right) \,=\left(
I\,+\,\Phi _{-}\right) ,  \tag{2.5}
\end{equation}
where\thinspace the integral operator $\Phi _F\left( \psi \right) \,$%
\thinspace is 
\begin{equation}
\Phi _F\left( \psi \right) \,=\,\int\limits_{-\infty }^\infty F\left(
t,z\right) \psi \left( z\right) dz.  \tag{2.6}
\end{equation}
Both $k_{+}\,$and$\,\,F\,\,$in eq.(2.4) and (2.6) are the 2x2 matrix chosen
as follows

\begin{equation}
k_{+}=\,\left( 
\begin{array}{cc}
a\left( t,z;x\right)  & q_1\left( t,z;x\right)  \\ 
\pm q_1^{*}\left( t,z;x\right)  & d\left( t,z;x\right) 
\end{array}
\right) ,  \tag{2.7a}
\end{equation}
and 
\begin{equation}
F\,=\,\left( 
\begin{array}{cc}
0 & \alpha _n^{\prime }\left( t,z;x\right)  \\ 
\beta _n^{\prime }\left( t,z;x\right)  & 0
\end{array}
\right) .  \tag{2.7b}
\end{equation}
Here $a$, $q_1$, $\pm q_1^{*}$, $d$, $\alpha _n^{\prime }$, and $\beta
_n^{\prime }$ are parameters which will be calculated in section III.

In eq.$(2.5)$, we have assumed that $\left( I\,+\,\Phi _{+}\right) \,\,$is
invertible, then 
\begin{equation}
\,\left( I\,+\,\Phi _F\right) \,=\left( I\,+\,\Phi _{+}\right) ^{-1}\left(
I\,+\,\Phi _{-}\right) ,  \tag{2.8}
\end{equation}
so that operator $\left( I\,+\,\Phi _F\right) \,$ is factorisable. From eq.$%
(2.5)$, we can derive Marchenko matrix equation, 
\begin{equation}
k_{+}\left( t,z\right) +\,F\left( t,z\right) \,+\,\int\limits_t^\infty
k_{+}\left( t,t^{\prime }\right) F\left( t^{\prime },z;x\right) dt^{\prime
}\,=\,0,\,\,\,\,\text{\thinspace \thinspace for}\,\,z\,>\,t\,\,  \tag{2.9a}
\end{equation}
and 
\begin{equation}
k_{-}\left( t,z\right) \,=\,F\left( t,z\right) \,+\,\int\limits_t^\infty
k_{+}\left( t,t^{\prime }\right) F\left( t^{\prime },z;x\right) dt^{\prime
}\,,\,\,\,\,\,\,\text{for}\,\,z\,<\,t.\,  \tag{2.9b}
\end{equation}
In eq.$(2.9b)$,\thinspace $k_{-}\,\,$is obviously defined in terms of $%
k_{+}\,$and\thinspace $F\,$.\thinspace Both eq.$(2.9a)\,$and $(2.9b)$%
\thinspace require $F$, and $F$ is supplied by the solution of equations : 
\begin{equation}
\left[ \Delta _0^{(1)},\Phi _F\right] =0,  \tag{2.10a}
\end{equation}
and 
\begin{equation}
\left[ \Delta _0^{(2)},\Phi _F\right] =0.  \tag{2.10b}
\end{equation}
After a little algebraic manipulation, we get 
\begin{equation}
\left( 
\begin{array}{cc}
m+1 & 0 \\ 
0 & -m
\end{array}
\right) F_t\,+\,F_z\left( 
\begin{array}{cc}
m+1 & 0 \\ 
0 & -m
\end{array}
\right) \,=\,0,\,  \tag{2.11a}
\end{equation}
and 
\begin{equation}
\frac i{\left( 2m+1\right) }F_x\,+\,F_{zz}\,-\,F_{tt}\,=\,0,\,\,\,\, 
\tag{2.11b}
\end{equation}
where $F_t\equiv \frac{\partial F}{\partial t}$, $F_z\equiv \frac{\partial F%
}{\partial z}$, etc.

$U\left( t,x\right) \,$and $V\left( t,x\right) \,$can be found by solving eq.%
$(2.3)\,$in which we have substituted eq.($2.7a$) and ($2.4$) (for $k_{+})\,$%
to that equation: 
\begin{equation}
V\left( t,x\right) \,=\,\left( 2m+1\right) \left( 
\begin{array}{cc}
0 & q_1 \\ 
\mp q_1^{*} & 0
\end{array}
\right) ,\,  \tag{2.12a}
\end{equation}
and 
\begin{equation}
U\left( t,x\right) \,=\,-2k_{+_t}\,=\,-2\left( 
\begin{array}{cc}
a_t & q_{1_t} \\ 
\pm q_{1_t}^{*} & d_t
\end{array}
\right) .  \tag{2.12b}
\end{equation}
Based on the solution of equation $\Delta ^{(2)}\left( I\,+\,\Phi
_{+}\right) \,=\,\left( I\,+\,\Phi _{+}\right) \Delta _0^{(2)}\,$,\thinspace
we get that $k_{+}\left( t,z;x\right) $ must obey the following equation: 
\begin{equation}
\left( 
\begin{array}{cc}
m+1 & 0 \\ 
0 & -m
\end{array}
\right) k_{+t}\,+\,k_{+z}\left( 
\begin{array}{cc}
m+1 & 0 \\ 
0 & -m
\end{array}
\right) +\,\left( 2m+1\right) \left( 
\begin{array}{cc}
0 & q_1 \\ 
\mp q_1^{*} & 0
\end{array}
\right) k_{+}\,=\,0\,,\,\,  \tag{2.13}
\end{equation}
and if we evaluate this eq.$(2.13)\,\,$on $z\,=\,t$, we find 
\begin{equation}
a_t\,=\,\mp \frac{\left( 2m+1\right) }{\left( m+1\right) }\mid q_1\mid
^2,\,\,\,\,\,\,d_t\,=\,\mp \frac{\left( 2m+1\right) }{\left( m\right) }\mid
q_1\mid ^2\,,  \tag{2.14a}
\end{equation}
\begin{equation}
\pm q_{1_t}^{*}\,=\,\mp \frac{\left( 2m+1\right) }{\left( m\right) }%
aq_1^{*},\,\,\,\,\,\,\,\,\,\,\,\,q_{1_t}\,=-\,\frac{\left( 2m+1\right) }{%
\left( m+1\right) }\,q_1d.\,\,  \tag{2.14b}
\end{equation}
Plugging the above equations into eq.$(2.12b)$ yields 
\begin{equation}
U\left( t,x\right) \,=\,-2\left( 
\begin{array}{cc}
\mp \frac{\left( 2m+1\right) }{\left( m+1\right) }\mid q_1\mid ^2 & q_{1_t}
\\ 
\pm q_{1_t}^{*} & \mp \frac{\left( 2m+1\right) }{\left( m\right) }\mid
q_1\mid ^2
\end{array}
\right) .  \tag{2.15}
\end{equation}
$\,\,\,\,\,\,\,\,\,\,\,\,\,\,\,\,\,\,\,\,\,\,\,\,\,\,\,\,\,\,\,\,\,\,\,\,\,%
\,\,\,\,\,\,\,\,\,\,\,\,\,\,\,\,\,\,\,\,\,\,\,\,\,\,\,\,\,\,\,\,\,\,\,\,\,\,%
\,\,\,\,\,\,\,\,\,\,\,\,\,\,\,\,\,\,\,\,\,\,\,\,\,\,\,\,\,\,\,\,\,\,\,\,\,\,%
\,\,\,\,\,\,$

Now we substitute equations $(2.12a)$ and $(2.15)$ into eq.$(2.2)$, we get 
\begin{equation}
\Delta ^{(1)}\,=\,I\left( \frac i{\left( 2m+1\right) }\frac \partial
{\partial x}\,-\,\frac{\partial ^2}{\partial t^2}\right) -2\left( 
\begin{array}{cc}
\mp \frac{\left( 2m+1\right) }{\left( m+1\right) }\mid q_1\mid ^2 & q_{1_t}
\\ 
\pm q_{1_t}^{*} & \mp \frac{\left( 2m+1\right) }{\left( m\right) }\mid
q_1\mid ^2
\end{array}
\right) ,  \tag{2.16a}
\end{equation}
and 
\begin{equation}
\Delta ^{(2)}\,=\,\left( 
\begin{array}{cc}
m+1 & 0 \\ 
0 & -m
\end{array}
\right) \frac \partial {\partial t}\,\,+\left( 2m+1\right) \left( 
\begin{array}{cc}
0 & q_1 \\ 
\mp q_1^{*} & 0
\end{array}
\right) \,.  \tag{2.16b}
\end{equation}
$\,\,\,\,\,\,$Since $\Delta ^{(1)}\,$commutes with $\Delta ^{(2)}$, we find 
\begin{eqnarray}
iq_{1_x}\,+\,q_{1_{tt}}\,\pm 2\left( \frac{\left( 2m+1\right) ^2}{-m}\,+\,%
\frac{\left( 2m+1\right) ^2}{\left( m+1\right) }\right) &\mid &q_1\mid
^2q_1\,=\,0  \nonumber \\
i\,q_{1_x}^{*}+\,q_{1_{tt}}^{*}\,\pm 2\left( \frac{\left( 2m+1\right) ^2}{-m}%
\,+\,\frac{\left( 2m+1\right) ^2}{\left( m+1\right) }\right) &\mid &q_1\mid
^2q_1^{*}\,=\,0.  \tag{2.17}
\end{eqnarray}
$\,$Eq.$(2.17)$ can then be manipulated as 
\begin{eqnarray}
iq_{1_x}\,+\,q_{1_{tt}}\,+2\mu \left( \mid q_1\mid ^2+\mid q_2\mid ^2\right)
q_1\, &=&\,0  \nonumber \\
i\,q_{1_x}^{*}+\,q_{1_{tt}}^{*}\,+2\mu \left( \mid q_1\mid ^2+\mid q_2\mid
^2\right) q_1^{*}\, &=&\,0,  \tag{2.18}
\end{eqnarray}
where we have used the relationship between $q_1\,\,$and$\,q_2\,$,\thinspace
namely 
\begin{equation}
q_2=\sqrt{2m}\,q_1^{*}\,,  \tag{2.19}
\end{equation}
and parameter $\mu \,$is an arbitrary positive real value

\begin{equation}
\mu \,=\,\frac{2m+1}{m^2\,+\,m}\,\,.  \tag{2.20}
\end{equation}

$\,\,\,\,$It is obvious that eq.$(2.18)$ is one part of the integrable
coupled nonlinear Schr\"{o}dinger equation of Manakov type (eq.$(1.1)$).$\,$

\section{The Bright \thinspace N-Solitons Solution}

We consider a general matrix function\thinspace \thinspace $F\,\,$in eq.($%
2.7b$)\thinspace and$\,$substitute\thinspace \thinspace it \thinspace into
eq.$(2.11b)$, we find two differential equations

\begin{equation}
\frac i{\left( 2m+1\right) }\alpha _{n_x}^{\prime }\,+\,\alpha
_{n_{zz}}^{\prime }\,-\,\alpha _{n_{tt}}^{\prime }\,=\,0,  \tag{3.1a}
\end{equation}
and 
\begin{equation}
\frac i{\left( 2m+1\right) }\beta _{n_x}^{\prime }\,+\,\beta
_{n_{zz}}^{\prime }\,-\,\beta _{n_{tt}}^{\prime }\,=\,0\,.  \tag{3.1b}
\end{equation}
$\,$The solution of the above$\,\,$equations$\,\,$can be derived by using
separable variable method. We then find

\begin{equation}
\alpha _n^{\prime }\left( t,z;x\right) \,=\,\sum\limits_{n=1}^N\alpha
_{n_0}^{\prime }e^{-\gamma _n\left( \left( m+1\right) z\right) }\left[
e^{-\gamma _n\left( mt-i\gamma _n\left( 2m+1\right) ^2x\right) }\right] , 
\tag{3.2a}
\end{equation}
and

\begin{equation}
\beta _n^{\prime }\left( t,z;x\right) \,=\,\sum\limits_{n=1}^N\beta
_{n_0}^{\prime }e^{\theta _n\left( mz\right) }\left[ e^{\theta _n\left(
\left( m+1\right) t-i\theta _n\left( 2m+1\right) ^2x\right) }\right] , 
\tag{3.2b}
\end{equation}
where $\alpha _{n_0}^{\prime }\,$\thinspace and\thinspace \thinspace $\beta
_{n_0}^{\prime }\,\,$are arbitrary constant parameters.\thinspace \thinspace 
$\,\,\,\,\,\,\,\,\,\,\,\,\,\,\,\,\,\,\,\,\,\,\,\,\,\,\,\,\,\,\,\,\,\,\,\,\,%
\,\,\,\,\,\,\,\,\,\,\,\,\,\,\,\,\,\,\,\,\,\,\,\,\,\,\,\,\,\,\,\,\,\,\,\,\,\,%
\,\,\,\,\,\,\,\,\,\,\,\,\,\,\,\,\,\,\,\,\,\,\,\,\,\,\,\,\,\,\,\,\,\,\,\,\,\,%
\,\,\,\,\,\,\,\,\,\,\,\,\,\,\,\,\,\,\,$

To get the final solution of the integrable coupled \textbf{NLS} equation of
Manakov type, we have to substitute eq.$(2.7b)$, eq.$(3.2a)\,$and$\,\,$eq.$%
(3.2b)\,\,$into Marchenko matrix equation (eq.$(2.9a)$). We get (for $a$ and 
$q_1$) 
\begin{equation}
a\left( t,z;x\right) \,=\,-\sum\limits_{n=1}^N\left( e^{m\theta _nz}\right)
\int\limits_t^\infty q_1\,\beta _{n_0}^{\prime }e^{\theta _n\left(
m+1\right) t^{\prime }}e^{-i\theta _n^2\left( 2m+1\right) ^2x}dt^{\prime
}\,\,,  \tag{3.3a}
\end{equation}
and 
\begin{eqnarray}
q_1\, &=&\,-\sum\limits_{n=1}^N\left( e^{-\left( m+1\right) \gamma
_nz}\right) \alpha _{n_0}^{\prime }e^{-m\gamma _nt}e^{i\left( 2m+1\right)
^2\gamma _n^2x}  \nonumber \\
&&\,-\sum\limits_{n=1}^N\left( e^{-\left( m+1\right) \gamma _nz}\right)
\int\limits_t^\infty a\left( t,z;x\right) \,\alpha _{n_0}^{\prime
}e^{-m\gamma _nt^{\prime }}e^{i\gamma _n^2\left( 2m+1\right)
^2x\,}dt^{\prime }\,.  \tag{3.3b}
\end{eqnarray}
Finally, by substituting eq.$(3.3a)$ to eq.$(3.3b)$ we find the solution : 
\begin{equation}
q_1\,=\,\sum\limits_{n=1}^N\frac{-\alpha _{n_0}^{\prime }\,e^{\left( -\left(
2m+1\right) \gamma _n\left[ t-i\left( 2m+1\right) \gamma _nx\right] \right)
}\,}{1\,+\,\left( \frac{-\mu \left( 2m+1\right) \alpha _{n_0}^{\prime }\beta
_{n_0}^{\prime }}{\left( -\left( 2m+1\right) \gamma _n\,+\,\left(
2m+1\right) \theta _n\right) ^2}\right) e^{-\left( 2m+1\right) \gamma
_n\left[ t-i\left( 2m+1\right) \gamma _nx\right] }e^{\left( 2m+1\right)
\theta _n\left[ t-i\left( 2m+1\right) \theta _nx\right] }}\,\,.  \tag{3.4}
\end{equation}
We define

\begin{equation}
\eta _n\,=\,k_n\left( t+ik_nx\right) ,  \tag{3.5a}
\end{equation}
and 
\begin{equation}
\eta _n^{*}\,=\,k_n^{*}\left( t-ik_n^{*}x\right) \,\,\,,\,  \tag{3.5b}
\end{equation}
where 
\begin{equation}
k_n\,=\,-\left( 2m+1\right) \gamma _n\,\,,\,  \tag{3.5c}
\end{equation}
and 
\begin{equation}
k_n^{*}\,=\,\left( 2m+1\right) \theta _n\,\,.  \tag{3.5d}
\end{equation}
Here $\,\gamma _n\,\,$and$\,\,\theta _n\,\,$are\thinspace
arbitrary\thinspace complex\thinspace parameters\thinspace and$\,-\gamma
_n^{*}\,=\theta _n.$

Now $q_1\,$(and $q_2$)\thinspace \thinspace can be rewritten as 
\begin{equation}
q_1\,=\sum\limits_{n=1}^N\frac{-\alpha _{n_0}^{\prime }\,e^{\eta _n}}{%
1\,+\,e^{R_n\,+\,\eta _n\,+\,\eta _n^{*}}}\,,  \tag{3.6}
\end{equation}
and 
\begin{equation}
q_2\,=\sum\limits_{n=1}^N\frac{-\sqrt{2m}\left( \alpha _{n_0}^{\prime
}\right) ^{*}\,e^{\eta _n^{*}}}{1\,+\,e^{R_n\,+\,\eta _n\,+\,\eta _n^{*}}}, 
\tag{3.7}
\end{equation}
where 
\begin{equation}
e^{R_n}\,=-\,\frac{\mu \left( 1+2m\right) \alpha _{n_0}^{\prime }\beta
_{n_0}^{\prime }}{\left( k_n\,+\,k_n^{*}\right) ^2}.  \tag{3.8}
\end{equation}
Based on our solution in eq.$(3.6)$ and $(3.7)$, we can see that our results
are the bright N-solitons since $\mu >0$. In the equations, there are only
two arbitrary complex parameters\thinspace ($\,\gamma _n\,\,$and$\,\,\theta
_n$)$\,$, and two arbitrary parameters $\alpha _{n_0}^{\prime }\,$and$%
\,\,\beta _{n_0}^{\prime }\,$which can directly influence the phase of the
solitons interaction. The results of $q_1$ (eq.$(3.6)$) and $q_2$ (eq.$(3.7)$%
) can be rewritten in the more conventional form by introducing $-\gamma
_n^{*}\,=-l_n+i\lambda _n$ (where $l_n$ and $\lambda _n$ are arbitrary real
parameter), 
\begin{equation}
q_1\,=\sum\limits_{n=1}^N\frac{\left( 2m+1\right) l_n\left( \frac{\alpha
_{n_0}^{\prime }}{\sqrt{\mu \left( 2m+1\right) \alpha _{n_0}^{\prime }\beta
_{n_0}^{\prime }}}\right) e^{\left[ -i\left( \left( 2m+1\right) \lambda
_nt\,+\,\left( 2m+1\right) ^2\left( \lambda _n^2-l_n^2\right) x\right)
\right] }\,}{\cosh \left[ -\left( 2m+1\right) l_nt\,-\,2\left( 2m+1\right)
^2l_n\lambda _nx\,+\,\varphi _n\right] },  \tag{3.9}
\end{equation}
and 
\begin{equation}
q_2\,=\sum\limits_{n=1}^N\frac{\left( 2m+1\right) \sqrt{2m}l_n\left( \frac{%
\left( \alpha _{n_0}^{\prime }\right) ^{*}}{\sqrt{\mu \left( 2m+1\right)
\left( \alpha _{n_0}^{\prime }\beta _{n_0}^{\prime }\right) ^{*}}}\right)
e^{\left[ i\left( \left( 2m+1\right) \lambda _nt\,+\,\left( 2m+1\right)
^2\left( \lambda _n^2-l_n^2\right) x\right) \right] }\,}{\cosh \left[
-\left( 2m+1\right) l_nt\,-\,2\left( 2m+1\right) ^2l_n\lambda
_nx\,+\,\varphi _n\right] }\,\,,  \tag{3.10}
\end{equation}
where $\varphi _n$ is a real N-solitons phase, 
\begin{equation}
\varphi _n\,=\,\frac 12R_n.\,\,  \tag{3.12}
\end{equation}

In our results, we have put a real phase $\varphi _n$ by replacing the
arbitrary negative parameters $\alpha _{n_0}^{\prime }\,$, the arbitrary
positive parameter $\,\beta _{n_0}^{\prime }$, and an arbitrary complex
parameter $k_n\,$which is related to arbitrary complex parameters $\gamma
_n\,\,$and$\,\,\theta _n$. The results derived above have contributed on the
obtaining a general form of an exact positive real parameter $\mu \,$and
have also shown that there is a special relationship (eq.$(2.19)$) between $%
q_1$ and $q_2$. Hence, by finding $q_1$,\thinspace we can directly get $q_2$%
\thinspace .

\thinspace \thinspace Based on our results in eq.$(3.6)$ and $(3.7)$, the
bright N-solitons solution can be derived into the bright one and two
soliton solutions related to\thinspace that in the works that have been done
before by Radhakrishnan, \thinspace et. \thinspace al.\thinspace \thinspace
using Hirota method in ref.$\left[ 8\right] $. According to the comparison
of both methods, their \thinspace results\thinspace of\thinspace the bright
one soliton solution\thinspace is equal to our results\thinspace when

$\alpha \,e^{\eta _1^{\left( 0\right) }}=\,-\alpha _{1_0}^{\prime }\,$,$%
\,\,\beta e^{\eta _1^{\left( 0\right) }}=-\sqrt{2m}\left( \alpha
_{1_0}^{\prime }\right) ^{*}e^{\eta _1^{*}}e^{-\eta _1}\,$,and$\,$

$\,\left( \mid \alpha \mid ^2+\mid \beta \mid ^2\right) =-\left( 1+2m\right)
\alpha _{1_0}^{\prime }\beta _{1_0}^{\prime }\,$.\thinspace

On the other hand,\thinspace their results of the inelastic\thinspace
collision of the bright two soliton\thinspace \thinspace can be reduced to
our\thinspace \thinspace elastic\thinspace \thinspace collision\thinspace
\thinspace of the solution if\thinspace we put $\alpha _1:\alpha _2=\beta
_1:\beta _2$\thinspace in\thinspace their\thinspace \thinspace result. So,
we can generalize that our result is the solution of an elastic collision of
the bright N-solitons.

\section{Discussions and Conclusions}

We have presented the bright N-solitons solution of the integrable coupled
NLS equation\thinspace of Manakov type using the N-solitons solution of the
integrable uncoupled NLS\thinspace equation based on Zakharov-Shabat
scheme.We can conclude that the solution of the Manakov type (where $\mu $
is exactly a positive real parameter) can be solved by using an expanded
inverse scattering Zakharov-Shabat scheme in which we have chosen a certain
operator in our solution (eq.$(2.1)$). The constraint or the relationship
between $q_1$ and\thinspace $q_2$ happens absolutely when parameter $\mu $
is equal to an exact positive real value shown in eq.$(2.20)$.

Finally, we find the novel result that the solution in eq.$(3.6)$ and $%
(3.7)\,\,$cor- responds to an elastic collision of the bright N-solitons, as
long as there is a constraint $q_2=\sqrt{2m}q_1^{*}$. From the eq.$(3.9)\,$%
and$\,$eq.$(3.10)$, we can also\thinspace conclude that although the
collision between the bright\thinspace N-solitons their\thinspace
velocities\thinspace and amplitudes or intensities do not change, their
phases do change.

According to the comparison between our one and two soliton solution of the
integrable coupled NLS equation of Manakov type and the solution provided by
Radhakrishnan,\thinspace R., et.al., we get that their inelastic\thinspace
collision of the\thinspace bright\thinspace two-soliton solution can be our
\thinspace elastic collision solution, \thinspace if some\thinspace certain
parameters in the results are omitted by putting $\alpha _1:\alpha _2=\beta
_1:\beta _2$.\thinspace In our results,\thinspace the parameter exp$\left(
\eta _j^{(0)}\right) $appeared in ref.[$8$] has been absorbed \thinspace
into $\alpha _{n_0}^{\prime }$. The explanations mentioned above mean that
there is an\thinspace N-solitons\thinspace solution of eq.$(1.1)$ in an
inelastic collision of the bright N-solitons if there is a
constraint\thinspace $q_2=\sqrt{2m}q_1^{*}$ , where $\mu \,$is an arbitrary
\thinspace positive real parameter. Investigations concerning this problem
are now in progress.

$\mathbf{Acknowledgements}$

Both authors would like to thank H.J.Wospakrik for useful discussions. We
also would like to thank P. Silaban for his encouragement. The work of F.Z.
is supported by the Hibah Bersaing VII/1, 1998-1999, DIKTI, Republic of
Indonesia.The work of H.E. is partially supported by PGSM scholarship,
Republic of Indonesia.

\end{document}